# Resilient Distribution System Restoration with Communication Recovery by Drone Small Cells

Haochen Zhang, *Student Member, IEEE*, Chen Chen, *Senior Member, IEEE,* Shunbo Lei, *Member, IEEE*, Zhaohong Bie, *Senior Member*

*Abstract*—Distribution system (DS) restoration after natural disasters often faces the challenge of communication failures to feeder automation (FA) facilities, resulting in prolonged load pick-up process. This letter discusses the utilization of drone small cells for wireless communication recovery of FA, and proposes an integrated DS restoration strategy with communication recovery. Demonstrative case studies are conducted to validate the proposed model, and its advantages are illustrated by comparing to benchmark strategies.

## I. Introduction

POST-DISASTER restoration is a critical measure in increasing the resilience of distribution system (DS). The disastrous events may not only damage the equipment of distribution grid to varying degrees, but also affect the communication infrastructure supporting monitor and control functions of feeder automation (FA). The cyber-physical interdependence will further complicate the restoration process and prolong the outage duration. In this sense, achieving the self-healing capability of communication network of FA in coordination with DS restoration is critical to enhance the system resilience.

Existing works on DS post-disaster restoration focus on facilitating supply continuity with distributed generations (DGs) and topology reconfiguration to form microgrids (MGs), e.g., [1] and [2]. This reconfiguration can be further coordinated with various flexible resources to boost restoration [3], [4]. However, current research usually neglects communication facility recovery of FA which supports fast restoration. On the other hand, the drone small cell (DSC) provides a promising solution for emergency communications. The DSC, which is an aerial wireless base station, can move in the affected area autonomously and provide a fast and reliable alternative for wireless communications. The utilization of DSCs in disaster recovery has demonstrated benefits, and some works further investigated optimal deployment strategies of DSCs for communication recovery [5] [6]. However, in DS restoration

H. Zhang, C. Chen and Z. Bie are with the State Key Laboratory of Electrical Insulation and Power Equipment, Shaanxi Province Key Laboratory of Smart Grid, Xi'an Jiaotong University, Xi'an 710049, China (e-mail: morningchen@xjtu.edu.cn).

S. Lei is with School of Science and Engineering, Chinese University of Hong Kong – Shenzhen, Guangdong 51872, China.

scenario, communication recovery of FA with DSCs should consider the cyber-physical interdependency, so the existing general communication recovery strategies for communication networks are not directly applicable.

To solve this issue, this letter develops an integrated DS restoration optimization model with communication recovery by DSCs. The solution can achieve optimal DSC deployment in coordination with DS restoration via MG formation, so as to pick up load as much as possible. In addition, we demonstrate the value of proposed integrated restoration strategy by comparing to general recovery strategy of communication system. To the best of our knowledge, this is the first work to integrate communication network recovery in the DS restoration strategy.

## II. Mathematical Formulation

### A. Communication Recovery Constraints by DSC

To accomplish speedy recovery of DS after a disaster, this letter adopts the resilient DS restoration model from [7], which uses a combination of different DGs and network reconfiguration strategies. The model fully utilizes topological flexibilities to form MGs energized by DGs through a set of innovative radiality constraints. We assume the feeder terminal units (FTUs) in FA, which control the automated feeder switches, communicate to the operation center via wireless cellular communication systems. By operating these switches, multiple MGs energized by DGs can be formed to quickly restore critical loads.

Due to the disaster, communication infrastructure (e.g., base stations) may be damaged so that FTUs will lose connections to operation center and the corresponding automated switches cannot be operated remotely for load restoration, even if the FTUs are intact and have the communication capability with backup power. In this situation, DSCs have the potential to set up emergency communications so that the intact FTUs can self-heal the communications to operation center to resume FA functions. The scarcity of DSCs requires optimal placement of drones, and the communication recovery should be coordinated with load restoration process. Specifically, the DS restoration model in [7] defines decision variables $\alpha_{ij}$ indicating status of branches $(i, j) \in L$ by operating switches. With the communication capability of switches by FTUs, $\alpha_{ij}$ depends not only on the operational and topological constraints of grid, but also on the DSC deployment strategy. Thus, the communication recovery constraints regarding the decision variable $\alpha_{ij}$ should be formulated.

This letter assumes communication coverage of DSC is approximated as a circle with fixed radius [6], representing the omnidirectional antenna of DSCs with fixed flying height. Considering there are switches at both ends of the branch, we use the directional arc $(i, j) \in S$ to index the switch close to feeder node $i$ on the branch $(i, j) \in L$. The switch $(i, j) \in S$ and the connected FTU are both co-located at the node $i$, so we use the longitude and latitude of the node to indicate the position of corresponding switch and FTU. Note that this setting also complies with the smart switches integrating the FTU function. The communication recovery constraints can be formulated as follows:

$$(x_{mn} - \bar{x}_k)^2 + (y_{mn} - \bar{y}_k)^2 = (d_{mn}^k)^2, \forall (m,n) \in S_f, \forall k \in U \quad (1)$$

$$\lambda_{mn}^k = \begin{cases} \{0,1\} & \text{if } d_{mn}^k \leq r \\ 0 & \text{otherwise} \end{cases} \forall (m,n) \in S_f, \forall k \in U \quad (2)$$

$$\sum_{(m,n) \in S_f} \lambda_{mn}^k \leq C_k, \forall k \in U \quad (3)$$

$$\gamma_{mn} = \begin{cases} 1 & \text{if } \lambda_{mn}^k = 1 \\ 0 & \text{otherwise} \end{cases} \forall (m,n) \in S_f, \forall k \in U \quad (4)$$

$$\gamma_{mn} = 1, \forall (m,n) \in S \setminus S_f \quad (5)$$

$$\mu_{ij} = \gamma_{ij} \cdot \gamma_{ji}, \forall (i,j) \in L \quad (6)$$

$$\alpha_{ij} = \begin{cases} \{0,1\} & \text{if } \mu_{ij} = 1 \\ L_{ij} & \text{Otherwise} \end{cases} \forall (i,j) \in L \quad (7)$$

where $U$ is the set of DSCs; $S_f$ is the set of switches that cannot be operated remotely due to the disaster; $x_{mn}$, $y_{mn}$, $\bar{x}_k$ and $\bar{y}_k$ are the longitude and latitude of switch $(m,n)$ and DSC $k$, respectively; $d_{mn}^k$ denotes the distance between switch $(m,n)$ and DSC $k$; $\lambda_{mn}^k \in \{0,1\}$ indicates whether the switching capability of switch $(m,n)$ is restored by DSC $k$, i.e., $\lambda_{mn}^k = 1$ if the switch is restored by DSC $k$, $\lambda_{mn}^k = 0$ otherwise; $C_k$ is the communication resource capacity of DSC $k$, i.e., the maximum number of FTUs it can connect; $\gamma_{mn} \in \{0,1\}$ indicates whether switch $(m,n)$ can be operated, i.e., if the switch can be operated, $\gamma_{mn} = 1$, otherwise, $\gamma_{mn} = 0$; $\mu_{ij} \in \{0,1\}$ indicates whether branch $(i, j)$ can be switched, i.e., $\mu_{ij} = 1$ if the branch can be switched, $\mu_{ij} = 0$ otherwise; $L_{ij} \in \{0,1\}$ expresses the initial state of branch $(i, j)$, i.e., if the branch is closed, $L_{ij} = 1$; otherwise, $L_{ij} = 0$.

Constraints (1)-(2) require the switches restored by DSC $k$ must be within its communication coverage. Due to the small area occupied by the DS, the distance $d_{mn}^k$ uses the Euclidean distance over a flat area instead of a sphere area. Constraint (1) can be relaxed as a conic constraint, and its feasibility region will be unchanged. Constraint (3) specifies the number of recovered switches as the communication capacity limitation of DSCs. Constraints (4)-(5) are to determine the remote controllability of all switches in DS, i.e., if the switch is in an area where communication is normal or in communication failure area but recovered by a DSC, the switch can be operated remotely. Constraint (6) means that the switching capability of a branch depends on whether the switches at both ends of the branch can be operated remotely. Constraint (7) describes the state of a branch that if the branch is controlled, it can be configured, otherwise it must keep the original state.

*B. Linear Reformulation of Model*

For the sake of model solving efficiency, we need to reformulate the nonlinear constraints (2), (4), (6), and (7) into linear ones. Constraints (2), (4), and (7) are judgment constraints and can be linearized in the similar way. As an example, consider the linearization of constraint (2), which can be linearized to constraint (8) by using the big-M approach.

$$(r - d_{mn}^k)/M - 1 \leq \lambda_{mn}^k \leq (r - d_{mn}^k)/M + 1, \\ \forall (m,n) \in S_f, \forall k \in U \quad (8)$$

where $M$ is a large enough positive number.

The analysis shows that: if $r \geq d_{mn}^k$, (8) can be schematized as $\varepsilon - 1 \leq \lambda_{mn}^k \leq \varepsilon + 1$, since $\lambda_{mn}^k$ is a binary variable, the value of $\lambda_{mn}^k$ can be set as 0 or 1; otherwise (8) corresponds to $-\varepsilon - 1 \leq \lambda_{mn}^k \leq -\varepsilon + 1$, which means $\lambda_{mn}^k$ can only set as 0. Similarly, constraints (4) and (7) can be reformulated to linear constraints (9)-(10):

$$\sum_{k \in U} \lambda_{mn}^k / M \leq \gamma_{mn} \leq \sum_{k \in U} \lambda_{mn}^k, \forall (m,n) \in S_f \quad (9)$$

$$(1 - \mu_{ij}) \cdot L_{ij} \leq \alpha_{ij} \leq \mu_{ij} + L_{ij}, \forall (i,j) \in L \quad (10)$$

Constraint (6) is nonlinear as a product of two binary variables, which can be reformulated to linear ones in (11).

$$\begin{cases} \mu_{ij} \leq \gamma_{ij} \\ \mu_{ij} \leq \gamma_{ji} \\ \gamma_{ij} + \gamma_{ji} - 1 \leq \mu_{ij} \end{cases} \forall (i,j) \in L \quad (11)$$

*C. Integrated DS Restoration Model with Communication Recovery*

For designing an integrated restoration strategy considering DS reconfiguration and DSC dispatch, a new formulation of integrated restoration model is constructed in (12), which is a mixed-integer second-order cone programming (MISOCP) model, and the objective function ensures the fundamental purpose is load recovery, on this premise, considering the shortest travel of all DSCs.

$$\max \sum_{i \in N} \delta_i \cdot p_i^c - \left( \sum_{k \in U} (\bar{x}_k - x_o)^2 + (\bar{y}_k - y_o)^2 \right) / M$$
$$\textbf{s.t.} \quad (1), (3), (5), (8)-(11) \quad (12)$$
$$(1)-(2), (11)-(21) \text{ in } [7]$$

where $N$ is the set of all DS nodes; $\delta_i \in \{0,1\}$ indicates if the load at node $i$ is picked up $\delta_i = 1$, and $\delta_i = 0$ otherwise; $p_i^c$ denotes active power demand of the load at node $i$; $x_o$ and $y_o$ are the initial position of DSCs.

The constraints (1)-(2) and (11)-(21) in [7] separately describes the topological and operational constraints, which ensures feasible operation and radiality of MG formation.

III. CASE STUDIES

The proposed model is validated on IEEE 33-node test feeder [8] with six DGs. There are 9 branches being open faults due to the disaster. In addition, the disaster destroys some base stations and thus causes communication failure of some areas in which 45 switches on 23 branches cannot be controlled unless recovered by DSCs. There are 3 DSCs pre-positioned near the substation, and their communication coverages are set the same. The optimization model is solved by Gurobi 9.5.0 with the default settings on a computer with an Intel i7-8700 processor and 8GB RAM.

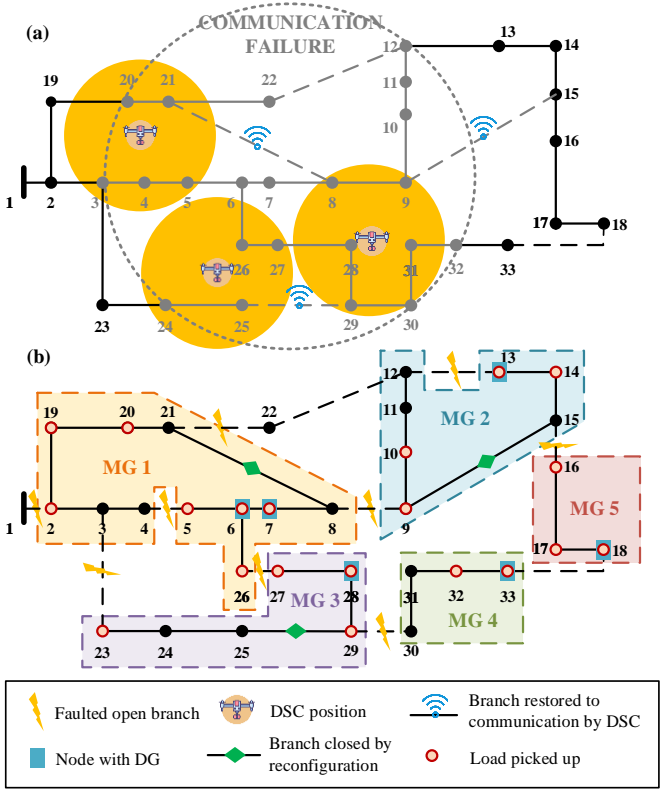

Fig. 1. Dispatching scheme of DSC and MG formation of DS.

TABLE I
RECOVERY RESULTS IN MULTIPLE STRATEGIES

| Restoration strategy | Restored load (MW) | Branch restored switching | Branch closed by reconfiguration |
|---|---|---|---|
| NoDSC | 46.2 | / | (18, 33) |
| Proposed model | 53.1 | (8, 21); (9, 15); (25, 29) | (8, 21); (9, 15); (25, 29) |
| MaxComm | 48 | (10, 11); (27, 28); (28, 29); (9, 15); (12, 22) | (9, 15); (18, 33) |

The results of the DSC dispatching strategy and the restoration decision of DS to form MGs are shown in Fig. 1. It can be seen that the DSCs are primarily used to recover switching capability of 3 branches and the DS is divided into 5 MGs by operating the switches, leading to a total of 53.1 MW load being picked up.

To verify the effectiveness of the proposed method, we design two benchmark strategies. One is using the restoration model in [7] without DSCs (denoted as NoDSC). The other is using the general communication recovery strategy, that is to maximize the number of users that can be recovered by DSC [5], and in DS, it means recovering the switching capability of more branches by using DSC (denoted as MaxComm). The comparisons of these three strategies are shown in Table I. We can see that for the NoDSC strategy, only branch (18, 33) is closed by reconfiguration and the restored load is only 46.2 MW on 16 feeder nodes in total. That means the disaster causes many FTUs to lose connection to the operation center due to the damage of base stations, resulting in a loss of partial FA functions and a significant degradation in DS recovery. In this sense, DSCs for restoring wireless communication of FA are critical to improve DS resilience.

Comparing the MaxComm strategy with the proposed strategy, there is also a reduction of load restoration, with loss of 5.1 MW restored load. Since in this strategy, the DSC concerns more on restoring switching capability of more branches, so the DSC dispatch position tends to be closer to the area where branches are more dense. Due to more scattered distribution of surrounding branches, the branches $(8, 21)$ and $(25, 29)$ cannot be switched, resulting in some nodes being isolated and unable to be energized by DG. Therefore, the general communication recovery strategy is not optimal since the cyber-physical interdependence between DS restoration and communication recovery is not considered. The integrated restoration strategy we proposed exploits the flexibility from the cyber-layer to achieve optimal decision of restoration from the physical-layer perspective.

## IV. CONCLUSION

This letter focuses on the utilization of DSCs to solve emergency communication needs of FA in DS post-disaster restoration. The communication recovery constraints of DSC deployment are formulated and integrated in DS restoration model. The use of DSCs can increase the topological flexibility from the cyber-layer perspective and the proposed restoration strategy with cyber-physical interdependence can enhance DS resilience in terms of load pickup when the communication infrastructure is impacted due to disasters.

REFERENCES

[1] C. Chen, J. Wang, F. Qiu and D. Zhao, "Resilient Distribution System by Microgrids Formation After Natural Disasters," *IEEE Trans. Smart Grid*, vol. 7, no. 2, pp. 958-966, March 2016.
[2] T. Ding, Y. Lin, G. Li and Z. Bie, "A New Model for Resilient Distribution Systems by Microgrids Formation," *IEEE Trans. Power Syst.*, vol. 32, no. 5, pp. 4145-4147, Sept. 2017.
[3] B. Li, Y. Chen, W. Wei, S. Huang and S. Mei, "Resilient Restoration of Distribution Systems in Coordination With Electric Bus Scheduling," *IEEE Trans. Smart Grid*, vol. 12, no. 4, pp. 3314-3325, July 2021.
[4] S. Yao, P. Wang and T. Zhao, "Transportable Energy Storage for More Resilient Distribution Systems With Multiple Microgrids," *IEEE Trans. Smart Grid*, vol. 10, no. 3, pp. 3331-3341, May 2019.
[5] P. Lohan and D. Mishra, "Utility-Aware Optimal Resource Allocation Protocol for UAV-Assisted Small Cells With Heterogeneous Coverage Demands," *IEEE Trans. Wirel. Commun.*, vol. 19, no. 2, pp. 1221-1236, Feb. 2020.
[6] M. Mozaffari, W. Saad, M. Bennis and M. Debbah, "Drone Small Cells in the Clouds: Design, Deployment and Performance Analysis," in *2015 IEEE Global Commun. Conf. (GLOBECOM)*, San Diego, CA, USA, 2015, pp. 1-6.
[7] S. Lei, C. Chen, Y. Song and Y. Hou, "Radiality Constraints for Resilient Reconfiguration of Distribution Systems: Formulation and Application to Microgrid Formation," *IEEE Trans. Smart Grid*, vol. 11, no. 5, pp. 3944-3956, Sept. 2020.
[8] M. E. Baran and F. F. Wu, "Network reconfiguration in distribution systems for loss reduction and load balancing," *IEEE Trans. Power Deliv.*, vol. 4, no. 2, pp. 1401-1407, April 1989.